\documentclass[journal,draftcls,onecolumn,12pt,twoside]{IEEEtranTCOM}

\usepackage{cite}
\usepackage{amsmath,amssymb,amsfonts}
\usepackage{algorithmic}
\usepackage{graphicx}
\usepackage{textcomp}
\usepackage{xcolor}
\usepackage{nccmath}
\usepackage{tikz}
\usepackage{textcomp}

\normalsize

\usepackage{cite}
\ifCLASSINFOpdf
\else
\fi

\begin{document}
\title{A Comprehensive Performance Analysis for mm-Wave Massive MIMO Hybrid Beamforming under PA Nonlinearities}

\author{Murat~Babek~Salman,~\IEEEmembership{Student~Member,~IEEE,}       
         Gokhan~Muzaffer~Guvensen,~\IEEEmembership{Member,~IEEE,}
        % and~Jane~Doe,~\IEEEmembership{Life~Fellow,~IEEE}% <-this % stops a space
%\thanks{M. B. Salman is with the Department
%of Electrical and Electronics Engineering, Middle East Technical University, Ankara, 06800 Turkey e-mail: mbsalman@metu.edu.tr.}% <-this % stops a space
\thanks{Murat~Babek~Salman and G. M. Guvensen are with the Department
of Electrical and Electronics Engineering, Middle East Technical University, Ankara, 06800 Turkey e-mail: mbsalman@metu.edu.tr, guvensen@metu.edu.tr.}% <-this % stops a space
}

\maketitle

\begin{abstract}
%\boldmath
In this paper, we develop a framework to investigate the performances of different hybrid beamforming architectures for massive multiple input multiple output (MIMO) systems impaired by power amplifier (PA) nonlinearities. Indirect learning architecture based on feedback after anti-beamforming is adopted in design of digital pre-distortion (DPD) in order to compensate for the nonlinear distortion caused by PA. In addition, we propose a novel analog beamformer design for partially connected array architecture based on generalized eigen-beamformer (GEB) approach. In literature, the effects of nonlinear PA's on the out-of-band (OOB) radiation and achieved signal-to-interference-plus-noise ratio (SINR) are investigated. However, these studies are limited to fully digital or partially connected hybrid beamforming architectures while deploying Bussgang decomposition on a PA basis without considering the array architecture type in performance analysis. In this study, we derived an analytical bit-error-rate (BER) expression based on spatio-frequency Bussgang decomposition in matrix form, and lower bound for mismatched decoding capacity via Generalized Mutual Information (GMI) is obtained under PA nonlinearity for different hybrid massive MIMO architectures. Analytical results show that the nonlinear distortion significantly affects the system performance, and DPD can reduce these effects to some extend. Finally, obtained analytical BER expression is verified via numerical results.
\end{abstract}


% \fi
\IEEEpeerreviewmaketitle

\section{Introduction}


\IEEEPARstart{M}{ultiple}-input multiple-output (MIMO) systems operating at milimeter-wave (mm-Wave) frequencies are expected to meet constantly growing data rate requirements that cannot be met by current systems \cite{mm_wave:1,mm_wave:2,mm_wave:3,mm_wave:4}. Radio signals, at mm-Wave frequencies, are exposed to severe absorption losses \cite{mm_wave:1,mm_wave:2}, hence, large scale antenna arrays, also called as massive MIMO systems, are used to increase the multiplexing gain so that effects of these losses are reduced \cite{mm_wave:3,mMIMO:1}.  However, using a large number of antennas has certain limitations. In conventional MIMO systems, which are employed for sub-6 GHz frequency bands, fully digital precoding can be implemented; however, doing so is quite costly for massive MIMO systems due to need of excessive number of RF chains, which consume immense amount of power \cite{FullDig,HybridBF:1,HybridBF:2}. %Massive MIMO transceivers are considered to be equipped with low-cost highly-efficient nonlinear power amplifiers (NPA)'s, which cause nonlinear distortion of transmitted signal, to reduce the hardware cost.

To reduce power consumption and implementation complexity, hybrid beamforming based array architectures have been developed. In these architectures, beamforming is divided into two stages, where precoding is performed in digital domain and analog beamformer connects $D$ RF chains to $N_t$ transmit antennas for $D<N_t$. One common approach is to implement analog beamformer as a partially connected structure, where each RF chain is connected to a subarray composed of a series of phase shifters \cite{HybridBF:3}, and digital precoder is designed based on the reduced dimensional effective channel. Joint spatial division and multiplexing (JSDM) is also proposed as a two stage beamforming framework \cite{JSDM} and \cite{JSDM2}. In JSDM, analog beamformer is designed based on spatial characteristics of user channels, which are characterized by their channel covariance matrices (CCM)'s, and users having similar CCM's are grouped so that they can be processed jointly. Analog beamformer is designed to eliminate inter-group interferences and decouples the signals of different groups. Digital precoder, on the other hand, is designed based on instantaneous channel in reduced dimension to suppress intra-group interference. In \cite{GEB:1} and \cite{GEB:2}, fully connected generalized eigen beamformer (GEB) is proposed as a statistical analog beamformer, which is shown to be optimal for several criteria. In proposed GEB, sub-beamformers are formed for each group such that beampattern contains deep nulls for angular sectors (AS)'s of other groups. This type of beamformer is suitable for the wideband massive MIMO channels, which are expected to be sparse both in the angle and delay domains \cite{JSDM2}.

In massive MIMO systems, low-cost PA's operating close to saturation are considered to be employed in future 5G systems in order to increase the power efficiency \cite{NonIdealArr}. It is observed that PA's exhibit nonlinear behavior around saturation region, which yields nonlinear distortion of transmitted signal. In this paper, we investigate the effects of PA nonlinearities on mm-wave Massive MIMO systems and examine compensation methods for such effects. 
\subsection{Related Literature}
Nonlinear distortion affects the system performance in several aspects, which are studied in literature \cite{OOBRad:1,OOBRad:2,SpatCharDist,DPD:1,DPD:2,DPD:3,AIR1:1,dinis_ber,abc,SER1,SER2,valkSINR}. In \cite{OOBRad:1,OOBRad:2,SpatCharDist}, out-of-band (OOB) radiation of massive MIMO systems are investigated with a particular focus on spatial characteristics of the radiation and a framework is developed for such analysis. These works consider fully digital beamforming and states that distortion effects are reduced as the number of antennas increases. Also, it was shown that as the number of users increases distortion power becomes isotropic and worst case happens, when a single user is present, where distortion is also concentrated towards that user. In \cite{DPD:1,DPD:2,DPD:3}, hybrid beamforming architecture is considered and both OOB radiation pattern and digital predistortion (DPD) techniques are studied. Different DPD design methods are proposed based on single PA feedback and anti-combining feedback architectures. It was shown that OOB radiation can be significantly reduced by DPD per RF chain. However, in \cite{MBS} we showed that DPD per RF chain cannot mitigate OOB radiation problem for fully connected hybrid architecture. Since, input of each PA is the combination of all RF chains, DPD per RF chain is not sufficient.

Furthermore, effects of nonlinear distortion on the capacity and error probability of the system are studied in  \cite{AIR1:1,dinis_ber,abc,SER1,SER2} and \cite{Capacity:1}. In  \cite{AIR1:1} and \cite{Capacity:1}, analysis on achievable information rate (AIR) is presented showing that capacity of massive MIMO systems is reduced due to nonlinear distortion. However, these studies consider the spectral efficiency in terms of unconstrained Shannon capacity and system is limited to be fully digital. In \cite{MBS}, we extended analysis to a more general framework where mismatched generalized mutual information (GMI) metric, which is  presented in \cite{Capacity:2,GMI:1,GMI:2,Zglgn}, is adopted to evaluate the performance of the system where fully connected hybrid architecture is considered. In literature, bit/symbol error probability (BER/SER) performances of fully digital massive MIMO systems are also studied by using narrowband Bussgang decomposition per antenna (scalar Bussgang with no frequency selectivity). In \cite{dinis_ber,abc,SER1,SER2}, both numerical and analytical BER/SER analysis are presented and it was shown that nonlinear distortion significantly increases BER of the system. %Besides, in \cite{valkSINR}, a signal-to-interference-plus-noise ratio (SINR) expression based on memory polynomial (MP) model is presented.

\subsection{Contributions}
In this paper, an extensive performance comparison between different massive MIMO array beamforming structures in JSDM framework under PA nonlinearity is carried out through complete analysis when proper DPD at base station (BS) or post-equalization at user side is utilized. A general analysis framework, covering different massive MIMO array types such as fully digital, fully connected hybrid/partially connected hybrid architectures, is proposed in order to quantify the performance losses in terms of radiation patterns, AIR, and BER under PA nonlinearities for multicarrier downlink transmission with higher order QAM constellations. To the author’s knowledge, in the recent literature, there is no such a comprehensive analytical investigation and comparison made before. 

Under the proposed framework, the contribution is two-fold. First, the nonlinearity due to PAs is modeled via vectorial wideband Bussgang decomposition in spatio-frequency domain, while taking the effect of hybrid beamforming structure into account. That is to say, the nonlinear system, from RF chain inputs including DPD units to PA outputs following the analog beamformer, is modeled in upsampled multidimensional signal domain. Here, the proposed modeling considers the spatial correlation among different transmit antennas at different subcarriers due to hybrid massive MIMO structure and PA memory. 

Second, based on the general spatio-frequency Bussgang decomposition, peculiar to hybrid array connection type, BER and AIR analysis are fulfilled for different conventional DPD techniques (based on memory polynomial model \cite{GMP}) and post frequency compensation methods at user terminal (UT). The effect of different beamforming types on the BER is demonstrated clearly both via the simulation and provided analysis. Moreover, mismatched decoding capacity via GMI is exploited to find a lower bound for AIR in case of mismatches (depending on array architecture) due to nonlinearity. It is observed that the hybrid array connection type is highly effective on the success of DPD and radiation patterns. 

As a side contribution, a novel GEB based partially connected subarray structure is proposed in order to compromise between the OOB pattern, nulling performance and BER. 

The remainder of this paper is organized as follows. In Section \ref{SysMod}, system and massive MIMO wideband channel models are presented. In Section \ref{TransSche}, transmission scheme and beamforming structures, which are employed in this study, are introduced. Section \ref{CompMeth} presents the compensation methods that are used to reduce the effects of PA nonlinearities. In Section \ref{PerfMeas}, performance measures, which are used in assessment of the performances, are presented and simulation results are given in Section \ref{NumRes}. Lastly, concluding remarks are stated in Section \ref{conclds}. Throughout the paper, scalars are notated with italic letters as $x$, vectors are notated with bold lower case letters as $\bf x$, matrices are notated with bold capital letters as $\bf X$. Conjugate transpose is denoted as $(\cdot)^H$, $\operatorname{Tr}(\cdot)$ denotes trace, and ${\bf X}[a,b]$ denotes the $(a,b)^{th}$ entry of matrix $\bf X$.

\section{System Model} \label{SysMod}

In the considered system, a base station, which is equipped with $\mathit{D} $ RF chains and $\mathit{N_{t}}$ antennas, serves $\mathit{U}$ single antenna users in time domain duplex (TDD) mode so that channel reciprocity can be exploited. Assuming a block fading frequency selective channel, which is slowly varying compared to signalling interval, received signal at the $\mathit{u^{th}}$ user can be given in time domain as
\begin{equation}
r^{({u})}_{n}=\sum_{l=0}^{L-1} [{\bf{h}}{_{l}^{({u})}}]{^H} {\bf{y}}_{n-l} + \nu^{({u})}_{n}, \label{rec_eq1}
\end{equation}
where, $\mathit{r_n^{({u})}}$ is the received signal, $\mathit{{\bf{h}}_{l}^{({u})}} \in \mathbb{C}^{N_{t} \times 1} $ is the $\mathit{l^{th}}$ delay of the channel impulse response of the $u^{th}$ user, $\mathit{{\bf{y}}_{n}}$  $\in \mathbb{C}^{N_{t} \times 1}$ is the transmitted signal, which is corrupted by the nonlinear PA, and $\mathit{\nu_n^{({u})}}$ is the additive white Gaussian noise (AWGN) with variance $N_o$.

\subsection{mm-Wave Wideband Massive MIMO Channel Model}
In this study, channel under consideration is taken to be sparse in both angular and temporal domains since massive MIMO systems are considered for millimeter (mm) wave frequencies, where channel sparsity is pronounced. By exploiting sparsity, multipath components stemming from the same angular sector can be grouped together so that JSDM can be used to reduce the effective dimension of the problem. In JSDM framework, users that reside in the same angular sector can be processed jointly since they share similar spatial statistics. Spatial characteristics of the multipath components (MPC)'s of each user of group $g$ are described by their CCM's, $\mathit{{\bf{R}}_{l}^{(g_u)}\in \mathbb{C}}^{N_t\times N_t} $, which can be expressed according to one-ring scattering model as \cite{JSDM2}
\begin{equation}
{\bf{R}}_l^{(g_u)} =\frac{{\gamma_l^{(g_u)}}}{2\Delta} \int_{\theta^{g_u,l}_c-\Delta}^{\theta^{g_u,l}_c+\Delta} {\bf{a}}(\theta)[{\bf{a}}(\theta)]^H d\theta,
\label{ccm_eq}
\end{equation}
%
%\begin{equation}
%\mathbb{E}\{{\bf{h}}_l^{(g_u)}[{\bf{h}}_l^{(g_u)}]^H\} = {\gamma_l^{(g)}} %{{\bf{R}}_l^{(g)}} {{\delta}_{gg^{'}}}{{\delta}_{uu^{'}}}{{\delta}_{ll^{'}}} ,%\label{ccm_eq}
%\end{equation}
%
where $ {\gamma_l^{(g_u)}} $ is the gain of the corresponding channel, which can also be inferred as power delay profile (pdp), and $ \operatorname{Tr}\{{\bf{R}}_l^{(g_u)}\}={\gamma_l^{(g_u)}}$,  $\theta^{g_u,l}_c$ is the center angle of arrival (AoA) for $l^{th}$ MPC of $u^{th}$ user of group $g$, and $ {\bf{a}}(\theta) \triangleq \frac{1}{\sqrt{N_t}} [1 \; e^{j\pi sin(\theta)} \; \ldots\; e^{j (N_t-1) \pi sin(\theta)}]^T$, is the unit norm steering vector for the azimuth angle $\theta$, which corresponds to uniform linear array (ULA) with half wavelength spacing. The spatial channel vector for $l^{th}$ MPC of the $u^{th}$ user in group $g$ is distributed as $ {\bf{h}}_l^{(g_u)} \sim \mathcal{C}\left({\kappa_l^{(g_u)}}{{\bf{a}}(\theta_c^{g_u,l}),{\bf{R}}_l^{(g_u)}} \right) $, where $\theta_c^{g_u,l}$ is the center AoA of the $l^{th}$ MPC of user $g_u$, ${\kappa_l^{(g_u)}}$ is a non-random complex constant with uniform phase distribution, and the channels of different users are independent. In the given channel model, it is assumed that each MPC is Rician distributed where a strong MPC exists at the center of the angular sector with Rician factor, $\left|{{\kappa^{(g_u)}_l}}\right|^2/{{\gamma_l^{(g_u)}}}$  \cite{ricianRef}. Note that, we can construct CCM for each group as, ${\bf R}_l^{(g)} \triangleq \sum_{u=1}^{U_g} {\bf R}_l^{(g_u)}$ \footnote{The design of user grouping algorithm is out of scope of this paper. An efficient procedure can be found in \cite{JSDM} and \cite{UserGr}.}. Then, received signal of the $u^{th}$ user of $g^{th}$ group can be expressed  as $r^{(g_{u})}_{n}=\sum_{l=0}^{L_g-1} \left[{\bf{h}}{_{l}^{(g_{u})}}\right]{^H} {\bf{y}}_{n-l} + \nu^{(g_{u})}_{n}$. Besides, it is assumed that instantaneous channel state information (CSI) is acquired perfectly during the uplink training.
%\begin{equation}
%r^{(g_{u})}_{n}=\sum_{l=0}^{L_g-1} [{\bf{h}}{_{l}^{(g_{u})}}]{^H} {\bf{y}}_{n-l} + \nu^{(g_{u})}_{n}. \label{rec_eq}
%\end{equation}
%

\subsection{Nonlinear PA Distortion}

In order to study the effects of the PA nonlinearities on the performance of the hybrid beamforming system, modelling the nonlinear PA is mandatory. Output of the nonlinear PA can be expressed in terms of arbitrary basis functions as

\begin{equation}
y^{(m)}_{n}=\sum_{\varpi=-\Pi+1}^{\Pi-1}  \sum_{\upsilon =0}^{\Upsilon-1} \alpha_{\varpi,\upsilon}^m\phi_{\varpi,\upsilon}(\bar{x}_{n-\varpi}^{(m)}),\label{pa_output}
\end{equation}
where $ y^{(m)}_n$ is the transmitted signal at the $ m^{th}$ antenna branch, $\phi_{\varpi,\upsilon}(\cdot)$ are the basis functions, $\bar{x}_{n-\varpi}^{(m)}$ is the input of nonlinearity and $\alpha$'s are the basis coefficients. The basis functions can be any function such as Memory Polynomial (MP) \cite{GMP} or Hermite Polynomials \cite{SpatCharDist} basis. In this work we consider MP expansion where the basis functions are $\phi_{\varpi,\upsilon}(\bar{x}_{n-\varpi})= {\bar{x}}(n-\varpi)|{\bar{x}}(n-\varpi)|^{2\upsilon}$.

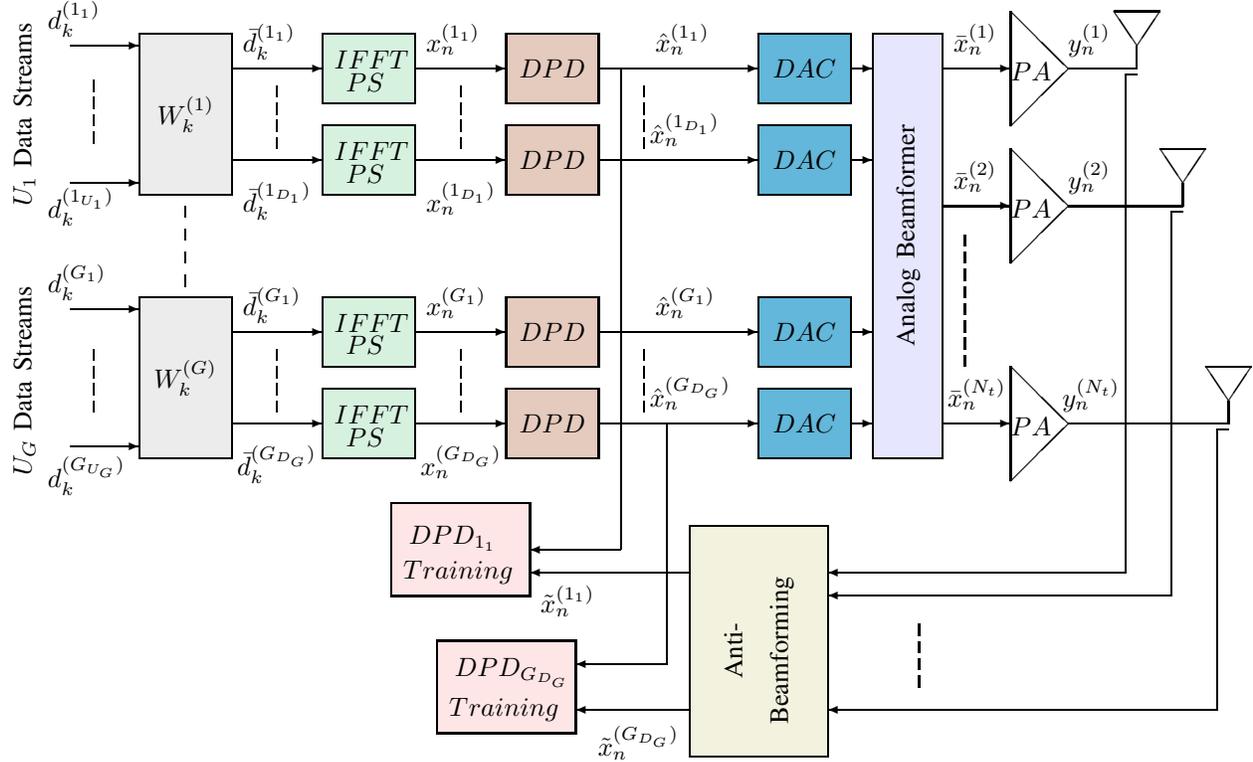
\begin{figure}[h]\setlength{\unitlength}{0.12in}
\centering% used for centering Figure
%\resizebox{80mm}{80mm}{
\begin{picture}(32,35)  
\setlength\fboxsep{0pt}
\small

\definecolor{mygreen}{RGB}{214, 242, 223}
\definecolor{mybrown}{RGB}{98, 183, 218}
\definecolor{dpdcol}{RGB}{229, 202, 190}

\put(-6,26.5){\colorbox{gray!15}{\framebox(4,7){$W_k^{(1)}$}}}
\put(2,30.5){\colorbox{mygreen}{\framebox(4,3){$ $}}}
\put(10,30.5){\colorbox{dpdcol}{\framebox(4,3){$DPD$}}}
\put(21,30.5){\colorbox{mybrown}{\framebox(4,3){$DAC$}}}
\put(2,26.5){\colorbox{mygreen}{\framebox(4,3){$ $}}}
\put(10,26.5){\colorbox{dpdcol}{\framebox(4,3){$DPD$}}}
\put(21,26.5){\colorbox{mybrown}{\framebox(4,3){$DAC$}}}

\put(2.5,32) {$IFFT$}
\put(3,31) {$PS$}

\put(2.5,28) {$IFFT$}
\put(3,27) {$PS$}

\put(-9,33){\vector(1,0){3}}
\multiput(-8,28.75)(0,0.75){4}{\line(0,1){0.5}}
\put(-9,27){\vector(1,0){3}}

\put(-2,32){\vector(1,0){4}}
\multiput(0,28.5)(0,0.75){4}{\line(0,1){0.5}}
\put(-2,28){\vector(1,0){4}}

\put(6,32){\vector(1,0){4}}
\multiput(8,28.5)(0,0.75){4}{\line(0,1){0.5}}
\put(6,28){\vector(1,0){4}}

\put(14,32){\vector(1,0){7}}
\multiput(16,28.5)(0,0.75){4}{\line(0,1){0.5}}
\put(14,28){\vector(1,0){7}}

\put(25,32){\vector(1,0){1}}
\put(25,28){\vector(1,0){1}}

\multiput(-4,22.5)(0,1){4}{\line(0,1){0.5}}

\put(-6,15){\colorbox{gray!15}{\framebox(4,7){$W_k^{(G)}$}}}
\put(2,19){\colorbox{mygreen}{\framebox(4,3){$ $}}}
\put(10,19){\colorbox{dpdcol}{\framebox(4,3){$DPD$}}}
\put(21,19){\colorbox{mybrown}{\framebox(4,3){$DAC$}}}
\put(2,15){\colorbox{mygreen}{\framebox(4,3){$ $}}}
\put(10,15){\colorbox{dpdcol}{\framebox(4,3){$DPD$}}}
\put(21,15){\colorbox{mybrown}{\framebox(4,3){$DAC$}}}

\put(2.5,20.5) {$IFFT$}
\put(3,19.55) {$PS$}

\put(2.5,16.5) {$IFFT$}
\put(3,15.5) {$PS$}

\put(-9,21.5){\vector(1,0){3}}
\multiput(-8,17)(0,0.75){4}{\line(0,1){0.5}}
\put(-9,15.5){\vector(1,0){3}}

\put(-2,20.5){\vector(1,0){4}}
\multiput(0,17)(0,0.75){4}{\line(0,1){0.5}}
\put(-2,16.5){\vector(1,0){4}}

\put(6,20.5){\vector(1,0){4}}
\multiput(8,17)(0,0.75){4}{\line(0,1){0.5}}
\put(6,16.5){\vector(1,0){4}}

\put(14,20.5){\vector(1,0){7}}
\multiput(16,17)(0,0.75){4}{\line(0,1){0.5}}
\put(14,16.5){\vector(1,0){7}}

\put(25,16.5){\vector(1,0){1}}
\put(25,20.5){\vector(1,0){1}}

\put(26,15){\colorbox{blue!10}{\framebox(3,18.5){$ $}}}
\put(27,19.5){\rotatebox{90}{ Analog Beamformer}}

\put(29,32){\vector(1,0){3}}
\put(29,26){\vector(1,0){3}}
\multiput(30,19)(0,0.75){8}{\line(0,1){0.5}}
\put(29,16.5){\vector(1,0){3}}

\put(32,29.5){\line(0,1){5}}
\put(32,34.5){\line(1,-1){2.5}}
\put(32,29.5){\line(1,1){2.5}}
\put(37,31.75){\line(1,0){0.5}}
\put(37,31.75){\line(0,-1){21.75}}
\put(37,10){\vector(-1,0){13}}
\put(34.5,32){\line(1,0){3}}
\put(37.5,32){\line(0,1){1}}	
\put(37.5,33){\line(2,3){1}}
\put(37.5,33){\line(-2,3){1}}
\put(36.5,34.5){\line(1,0){2}}
\put(32,31.5) {$PA$}

\put(32,23.5){\line(0,1){5}}
\put(32,28.5){\line(1,-1){2.5}}
\put(32,23.5){\line(1,1){2.5}}
\put(39,25.75){\line(1,0){0.5}}
\put(39,25.75){\line(0,-1){16.75}}
\put(39,9){\vector(-1,0){15}}
\put(34.5,26){\line(1,0){5}}
\put(39.5,26){\line(0,1){1}}	
\put(39.5,27){\line(2,3){1}}
\put(39.5,27){\line(-2,3){1}}	
\put(38.5,28.5){\line(1,0){2}}
\put(32,25.5) {$PA$}

\put(32,14){\line(0,1){5}}
\put(32,19){\line(1,-1){2.5}}
\put(32,14){\line(1,1){2.5}}
\put(41,16.25){\line(1,0){0.5}}
\put(41,16.25){\line(0,-1){12.25}}
\put(41,4){\vector(-1,0){17}}
\put(34.5,16.5){\line(1,0){7}}	
\put(41.5,16.5){\line(0,1){1}}	
\put(41.5,17.5){\line(2,3){1}}
\put(41.5,17.5){\line(-2,3){1}}	
\put(40.5,19){\line(1,0){2}}
\put(32,16) {$PA$}

\multiput(28,5)(0,0.75){4}{\line(0,1){0.5}}

\definecolor{antibeam}{RGB}{243, 242, 223}
\put(18,2){\colorbox{antibeam}{\framebox(6,10){$ $}}}
\put(19.5,5){\rotatebox{90}{ Anti-}}
\put(21.5,3){\rotatebox{90}{ Beamforming}}

\put(18,10){\vector(-1,0){7}}
\put(5,9){\colorbox{red!10}{\framebox(6,4){$ $}}}
\put(15,32){\line(0,-1){21}}
\put(15,11){\vector(-1,0){4}}
\put(5.75,11.25) {$DPD_{{1_1}}$}
\put(5.5,9.75) {$Training$}
\put(11.5,8.25) {${\tilde{x}}_n^{(1_1)}$}

\put(18,4){\vector(-1,0){5}}
\put(7,3){\colorbox{red!10}{\framebox(6,4){$ $}}}
\put(17,16.5){\line(0,-1){10.5}}
\put(17,6){\vector(-1,0){4}}
\put(7.5,3.75) {$Training$}
\put(7.75,5.5) {$DPD_{G_{D_G}}$}
\put(14,2) {${\tilde{x}}_n^{(G_{D_G})}$}

\put(-10,33.75) {$d_k^{(1_1)}$}
\put(-10,25.5) {$d_k^{(1_{U_1})}$}
\put(-11.5,26){\rotatebox{90}{\small $U_1$ Data Streams}}

\put(-10,22.25) {$d_k^{(G_1)}$}
\put(-10,13.75) {$d_k^{(G_{U_G})}$}
\put(-11.5,14.25){\rotatebox{90}{\small $U_G$ Data Streams}}

\put(-1.5,32.75) {${\bar{d}}_k^{(1_1)}$}
\put(-1.5,25.75) {${\bar{d}}_k^{(1_{D_1})}$}

\put(-1.5,21.25) {${\bar{d}}_k^{(G_1)}$}
\put(-1.75,14.25) {${\bar{d}}_k^{(G_{D_G})}$}

\put(6.5,32.75) {$x_n^{(1_1)}$}
\put(6.5,25.75) {$x_n^{(1_{D_1})}$}

\put(6.5,21.25) {$x_n^{(G_1)}$}
\put(6.25,14.25) {$x_n^{(G_{D_G})}$}

\put(16.5,21.25) {$\hat{x}_n^{(G_1)}$}
\put(16.25,17.25) {$\hat{x}_n^{(G_{D_G})}$}

\put(16.5,32.75) {$\hat{x}_n^{(1_1)}$}
\put(16.25,28.75) {$\hat{x}_n^{(1_{D_1})}$}

\put(29.5,32.75) {${\bar{x}}_n^{(1)}$}
\put(29.5,26.75) {${\bar{x}}_n^{(2)}$}
\put(29.25,17.25) {${\bar{x}}_n^{(N_t)}$}

\put(34.5,32.75) {$y_n^{(1)}$}
\put(34.5,26.75) {$y_n^{(2)}$}
\put(34.25,17.25) {$y_n^{(N_t)}$}
\end{picture}
%}
\caption{Hybrid beamforming system architecture. For the sake of simplicity continuous time variables, $\bar{x}$ and $y$, are denoted by upsampled discrete time time index, $n$}% title of the Figure
\label{sys_arch}% label to refer figure in text
\end{figure}

\section{A Generic Transmission Scheme} \label{TransSche}
A generic transmitter structure based on hybrid bemforming architecture is shown in Fig. \ref{sys_arch}. In the considered structure, $ U $ users are grouped into $ G $ groups, each having $U_g$ users. In addition, $ D_g$ RF chains are assigned to each group, $ \sum_{g=1}^{G} D_g = D$, where $D<N_t$ is the total number of available RF chains. Main motivation behind the use of hybrid beamforming is that reduced number of RF chains is sufficient in hybrid structure as opposed to fully digital structure, which requires as many RF chains as the number of antennas. Even though, the considered structure defines a hybrid beamforming architecture, it is easily extended to fully digital architecture as $ G=1$ and $D = N_t$. Transmitted upsampled signal can be formed by utilizing 
Orthogonal frequency-division multiplexing (OFDM) modulation as
\begin{equation}
{\bf \bar{x}}_n=\frac{1}{\sqrt{K}}\sum_{k=0}^{ \mu K-1} {\bf{B}}{\bf{W}}_k {\bf{d}}_k e^{j\frac{2\pi}{\mu K}kn}, \label{tx_sig_eq}
\end{equation}
where $K$ is the number of QAM modulated active subcarriers, $\mu$ is the oversampling factor, $ {\bf B}\in {\mathbb{C}}^{N_t\times D} $ is the analog beamformer matrix, $ {\bf W}_k\in {\mathbb{C}}^{D\times U} $ is the digital precoder matrix for $k^{th}$ subcarrier, $ \mathbb{\bf{d}}_k = [d_k^1, d_k^2, ..., d_k^{U}]^T \in \mathbb{C}^{U \times 1}$ is composed of $\mathit{i.i.d.}$ QAM modulated data symbols with autocorrelation, $ \mathit{\mathbb{E}\{{\bf{d}}_k{\bf{d}}_{k-l}^H\} = {\bf{I}} \delta_l}$, and constellation order $M$. 

Due to impracticality of joint design of digital and analog beamformers, JSDM framework is adopted. In this framework, designs of beamformers are decoupled such that analog beamformer is formed based on second order statistics of the channels, while the digital precoder is designed by using reduced rank instantaneous effective channel. Digital precoder constitutes the first stage of the beamforming, which is used to isolate the intra-group users. 

\subsection{Design of Digital Precoder}
In this work, well-known zero-forcing (ZF) precoding is employed, which is designed based on the effective channel matrix of each group, $\mathit{{\bf{H}}_{eff,l}^{(g)}} \triangleq [{\bf{B}}^{(g)}]^H{\bf{H}}_{l}^{(g)}$ ,where $ {\bf{B}}^{(g)} \in \mathbb{C}^{N_t \times D_g }$ for $ g = 1,...,G $ is the sub-beamforming matrix for group $ g $ and $ {\bf{H}}_{l}^{(g)} = [{\bf{h}}_{l}^{(g_1)}, ..., {\bf{h}}_{l}^{(g_{U_g})} ]$ is the full dimension channel matrix for group $g$. Analog beamforming matrix is formed as the combination of the sub-beamformers for each group as, $ {\bf{B}} = [{\bf{B}}^{(1)}, {\bf{B}}^{(2)}, ..., {\bf{B}}^{(G)}]$. In general, channel estimation is performed in reduced dimension; however, we assume that full rank channel is available at the BS since channel estimation is not in our scope. Frequency domain regularized ZF precoder with regularization parameter $\delta$ for a fixed analog beamformer is obtained as
\begin{equation}
{\bf {W}}_k^{(g)} \triangleq {{\bf{\Omega}}^{(g)}_{eff,k}}\left([{{\bf{\Omega}}^{(g)}_{eff,k}}]^H {{{\bf{\Omega}}^{(g)}_{eff,k}}} + \delta {\bf I}_{U_g}\right)^{-1},\label{ZF_ofdm} 
\end{equation}
where ${\bf{\Omega}}^{(g)}_{eff,k}= [{\bf{B}}^{(g)}]^H {\bf{\Omega}}^{(g)}_k$ is effective channel matrix in frequency domain and $ {\bf{\Omega}}_{k}^{(g)} = \sum_{l=0}^{\mu K-1}{\bf{H}}{_{l}^{(g)}}e^{-j2\pi l \frac{k}{\mu K}} $ is the frequency domain response for subcarrier $k $ of group $ g $.

Overall digital precoding matrix has the block diagonal form ${\bf{W}}_k \triangleq  \operatorname{bdiag}\{ \sqrt{c^{(g)}} {\bf{W}}_{k}^{(g)} \}_{g=1}^{G}$, where $c^{(g)} $ is the power scaling factor, then digitally precoded upsampled transmitted signal can be expressed as, ${\bf \bar{x}}_n=\frac{1}{\sqrt{K}}\sum_{k=0}^{ \mu K-1} \sum_{g=1}^{ G} \sqrt{c^{(g)}} {\bf{B}}^{(g)}{\bf{W}}^{(g)}_k {\bf{d}}_k^{(g)} e^{j\frac{2\pi}{\mu K}kn}$, where ${\bf d}_k^{(g)} \triangleq [d_k^{(g_1)}\; \ldots \; d_k^{(g_{U_g})}]^T$. In the considered scenario, per group power constraint is adopted to obtain  $c^{(g)}  = E_s/(G P^{(g)}) $, where $E_s = \mathbb{E}\{ |{\bf \bar{x}}_n|^2 \}$ is the total transmit power, and ${ P^{(g)}} $ is
\begin{equation}
{ P^{(g)}} \triangleq  \frac{1}{K} \sum_{k=0}^{K-1} \operatorname{Tr}\left\{ {\bf B}^{(g)} {\bf{W}}_{k}^{(g)} \left[{\bf{W}}_{k}^{(g)}\right]^H \left[{\bf B}^{(g)}\right]^H\right\}. 
\label{pow_sc}
\end{equation}
Based on the power constraint expression in \eqref{pow_sc}, it can be inferred that transmit power is kept constant for each channel realization.

\subsection{Design of Analog Beamformer}
In hybrid structures, analog beamforming is applied to radiate the energy in the direction of the group of the intended user. Analog beamformers can be classified into two groups based on the antenna array structure, namely "fully connected" and "partially connected" arrays. In fully connected array architecture, shown in Fig. \ref{array_arch}(a), each antenna input is combination of all RF chains, while in partially connected array architecture, shown in Fig. \ref{array_arch}(b), each antenna input is dedicated to a single RF chain.

%\begin{figure}
%\centering
   % \includegraphics[scale=0.9]{array_arch.eps}
    %\caption{ Analog beamforming architecutres (a) Fully connected (b) Partially connected architecure.}
    %\label{array_arch}
%\end{figure}

\usetikzlibrary{shapes,arrows}

\tikzset{%
  block/.style    = {draw, thick, rectangle, minimum height = 3cm,
    minimum width = 3em},
  sum/.style      = {draw, circle, node distance = 2cm}, % Adder
  sumq/.style      = {draw, circle, node distance = 2cm}, % Adder
  gain/.style      = {draw, circle, node distance = 2cm}, % Adder
  input/.style    = {coordinate}, % Input
  output/.style   = {coordinate} % Output
}

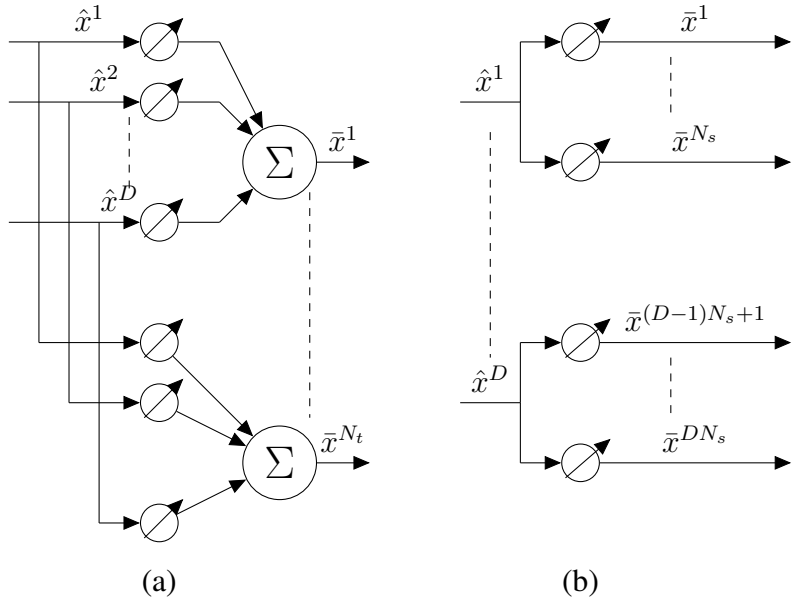
\begin{figure}[h]\setlength{\unitlength}{0.12in}
\centering% used for centering Figure
\newcommand{\suma}{\Large$\Sigma$}
\newcommand{\sumaq}{\Large$\Sigma$}
\newcommand{\inte}{$\displaystyle \int$}
\newcommand{\derv}{\huge$\frac{d}{dt}$}
\begin{tikzpicture}[auto, node distance=2cm, >=triangle 45, scale=0.8]

\draw

	node [input, name=input1] at (-9.5,1) {} 
	node [input, name=input2] at (-9.5,0) {} 
	node [input, name=input3] at (-9.5,-2) {} 

	node [input, name=input11] at (-9,1) {} 
	node [input, name=input21] at (-8.5,0) {} 
	node [input, name=inputN1] at (-8,-2) {} 

	node [input, name=input12] at (-6,1) {} 
	node [input, name=input22] at (-6,0) {} 
	node [input, name=inputN2] at (-6,-2) {} 

	node [input, name=input1D] at (-9,-4) {} 
	node [input, name=input2D] at (-8.5,-5) {} 
	node [input, name=inputND] at (-8,-7) {} 

	node [input, name=p11] at (-7.25,0.75) {} 
	node [input, name=p12] at (-6.6,1.4) {} 

	node [input, name=p21] at (-7.25, -0.25) {} 
	node [input, name=p22] at (-6.6, 0.4) {} 

	node [input, name=pN1] at (-7.25,-2.25) {} 
	node [input, name=pN2] at (-6.6,-1.6) {}

	node [input, name=pD1] at (-7.25,-4.25) {} 
	node [input, name=pD2] at (-6.6,-3.6) {} 

	node [input, name=pD3] at (-7.25, -5.25) {} 
	node [input, name=pD4] at (-6.6, -4.6) {} 

	node [input, name=pD5] at (-7.25,-7.25) {} 
	node [input, name=pD6] at (-6.6,-6.6) {}

	node [sum] (suma1)  at (-5,-1) {\suma}
	node [sumq] (suma2)  at (-5,-6) {\sumaq}

	node [input, name = xsum1] at (-3.5,-1) {} 
	node [input, name = xsumN] at (-3.5,-6) {};

\node [circle ] (labela) at (-7,-8) {(a)} ;
\draw[-](input1) -- node{ }(input11);
\draw[-](input2) -- node{ }(input21);
\draw[-](input3) -- node{ }(inputN1);

\draw [dashed] (-4.5,-1.5) -- (-4.5,-5.25);

\draw[->](suma1) -- node{ $\bar{x}^1$}(xsum1);
\draw[->](suma2) -- node{ $\bar{x}^{N_t}$}(xsumN);

\node [circle , minimum height = 0.5cm ,draw=black] (p1) at (-7,1) {};
\draw[->](input11) -- node{$ \hat{x}^1$} (p1);
\draw[->](p11) -- node {}(p12);
\draw[-](p1) -- node{ }(input12);
\draw[->](input12) -- node{ }(suma1);

\node [circle , minimum height = 0.5cm ,draw=black] (p2) at (-7, 0) {};
\draw[->](input21) -- node{$ \hat{x}^2$} (p2);
\draw[->](p21) -- node {}(p22);
\draw[-](p2) -- node{ }(input22);
\draw[->](input22) -- node{ }(suma1);

\draw [dashed] (-7.5,-0.25) -- (-7.5,-1.5);

\node [circle , minimum height = 0.5cm ,draw=black] (pN) at (-7,-2) {};
\draw[->](inputN1) -- node{$ \hat{x}^D$} (pN);
\draw[->](pN1) -- node {}(pN2);
\draw[-](pN) -- node{ }(inputN2);
\draw[->](inputN2) -- node{ }(suma1);

\node [circle , minimum height = 0.5cm ,draw=black] (p1N) at (-7,-4) {};
\draw[->](pD1) -- node {}(pD2);
\draw[-](input11) --  node{} (input1D);
\draw[->](input1D) --  node{} (p1N);
\draw[->](p1N) -- node{ }(suma2);

\node [circle , minimum height = 0.5cm ,draw=black] (p2N) at (-7,-5) {};
\draw[->](pD3) -- node {}(pD4);
\draw[-](input21) --  node{} (input2D);
\draw[->](input2D) --  node{} (p2N);
\draw[->](p2N) -- node{ }(suma2);

\node [circle , minimum height = 0.5cm ,draw=black] (pDN) at (-7,-7) {};
\draw[->](pD5) -- node {}(pD6);
\draw[-](inputN1) --  node{} (inputND);
\draw[->](inputND) --  node{} (pDN);
\draw[->](pDN) -- node{ }(suma2);

%%%%%%%%%%%%%%%%%%%%%%%%%%%%%%%%%%%%
% Partially Connected
%%%%%%%%%%%%%%%%%%%%%%%%%%%%%%%%%%%%

\node [input, name=part11] at (-2,0) {}; 
\node [input, name=part12] at (-1,0) {};
\node [input, name=part1u] at (-1,1) {};
\node [input, name=part1d] at (-1,-1) {};

\draw [dashed] (1.5,0.75) -- (1.5,-0.25);
\draw [dashed] (1.5,-4.25) -- (1.5,-5.25);

\node [input, name=ph11] at (-0.25,0.75) {} ;
\node [input, name=ph12] at (0.5,1.4) {} ;

\node [input, name=ph21] at (-0.25,-1.25) {} ;
\node [input, name=ph22] at (0.5,-0.6) {} ;

\node [input, name=part21] at (-2,-5) {}; 
\node [input, name=part22] at (-1,-5) {};
\node [input, name=part2u] at (-1,-4) {};
\node [input, name=part2d] at (-1,-6) {};

\node [input, name=ph31] at (-0.25,-4.25) {} ;
\node [input, name=ph32] at (0.5,-3.6) {} ;

\node [input, name=ph41] at (-0.25,-6.25) {} ;
\node [input, name=ph42] at (0.5,-5.6) {} ;

\node [input, name=op1] at (3.5,1) {}; 
\node [input, name=op2] at (3.5,-1) {};
\node [input, name=op3] at (3.5,-4) {};
\node [input, name=op4] at (3.5,-6) {};

\node [circle ] (labela) at (-0,-8) {(b)} ;

\draw [dashed] (-1.5,-0.5) -- (-1.5,-4.25);

\draw[-](part11) -- node {$ \hat{x}^1$}(part12);
\draw[-](part12) -- node {}(part1u);
\draw[-](part12) -- node {}(part1d);
\node [circle , minimum height = 0.5cm ,draw=black] (ph1) at (0,1) {};
\draw[->](ph11) -- node {}(ph12);
\draw[->](part1u) -- node {}(ph1);
\node [circle , minimum height = 0.5cm ,draw=black] (ph2) at (0,-1) {};
\draw[->](ph21) -- node {}(ph22);
\draw[->](part1d) -- node {}(ph2);

\draw[-](part21) -- node {$ \hat{x}^D$}(part22);
\draw[-](part22) -- node {}(part2u);
\draw[-](part22) -- node {}(part2d);
\node [circle , minimum height = 0.5cm ,draw=black] (ph3) at (0,-4) {};
\draw[->](ph31) -- node {}(ph32);
\draw[->](part2u) -- node {}(ph3);
\node [circle , minimum height = 0.5cm ,draw=black] (ph4) at (0,-6) {};
\draw[->](ph41) -- node {}(ph42);
\draw[->](part2d) -- node {}(ph4);

\draw[->](ph1) -- node {$\bar{x}^1$}(op1);
\draw[->](ph2) -- node {$\bar{x}^{N_s}$}(op2);
\draw[->](ph3) -- node {$\bar{x}^{(D-1)N_s+1}$}(op3);
\draw[->](ph4) -- node {$\bar{x}^{D N_s}$}(op4);

\end{tikzpicture}
\caption{Analog beamforming architecutres (a) fully connected (b) partially connected architecure}% title of the Figure
\label{array_arch}% label to refer figure in text
\end{figure}

\subsubsection{Fully Connected Array}

For this architecture, angle only generalized eigen beamformer (AO-GEB), which is shown to be nearly optimal for several criteria in \cite{GEB:1}, is employed. Due to frequency domain equalization performed in OFDM modulation, CCM of the frequency domain response should be used in design of analog beamformer. Since each MPC of a group is independent, we can show that $  \mathbb{E}\{ {\bf{\Omega}}_{k}^{(g)} [{\bf{\Omega}}_{k}^{(g)}]^H\} = \sum_{l=0}^{L_g} {\bf{R}}_{l}^{(g)}$. Hence, summation of CCM's of MPC's is used in GEB design. For each group, a sub-beamformer is formed as the first dominant $D_g$ eigenvectors of the generalized eigenvalue problem \cite{GEB:1}
\begin{equation}
{\bf B}^{(g)} = \operatorname{eigs}({\bf R}_{sum}^{(g)},{\bf R}_{sum},D_g), \label{ao_geb}
\end{equation}
where ${\bf{R}}_{sum}^{(g)} \triangleq  \frac{E_s }{G} \sum_{l=0}^{L_g-1}{\bf{R}}^{(g)}_l$ and ${\bf{R}}_{sum} \triangleq \frac{E_s}{G}\sum_{g=1}^{G} \sum_{l=0}^{L_g-1}{\bf{R}}^{(g)}_l + N_o \bf I$. In GEB, objective is to maximize the radiated power in the direction of the intended group while minimizing that of the other groups, and GEB here is a kind of statistical Capon beamformer for general-rank signal models \cite{GEB:3}. Despite the fact that fully connected array architecture performs close to fully digital beamforming, it suffers from high complexity of implementation. 

\subsubsection{Partially Connected Array}

Partially connected array architecture is a feasible structure to reduce the complexity of the array. For this architecture, analog beamforming matrix has the block diagonal form ${\bf B}^{(g)} = \operatorname{bdiag}\{{\bf b}_1^{(g)} , {\bf b}_2^{(g)}, ...,{\bf b}_{D_g}^{(g)}\}$, where $ {\bf b}_d^{(g)} = [b_d^1,b_d^2,...,b_d^{N_s}]^T \in \mathbb{C}^{N_s \times 1 }$ is the sub-beamformer for the antenna array with $N_s$ antennas of $ d^{th}$ RF chain and ${\bf B} = \operatorname{bdiag} \{{\bf B}^{(1)}, \; \ldots, \; {\bf B}^{(G)}\}$. Phase only beamforming is a common approach, in which beamforming is applied via phase shifters where  $|b_d^{m}| =1$ $ \forall d,m $. In this work, DFT beamformer is employed for phase only subarray, which steers toward the AoA of the strong MPC of each user without considering inter-group interference \cite{GEB:2}.

%\begin{equation}
%{\bf B} = bdiag({\bf b}_1 , {\bf b}_2, ...,{\bf b}_D), \label{B_mat}
%\end{equation}

In addition to phase only subarray, we propose a novel sub-beamformer design method for partially connected array architecture based on generalized eigen beamforming. Since phase only beamformers cannot suppress the inter-group interference sufficiently, we adopt the GEB design, which is shown to be optimal for fully connected array architecture. In the proposed method, each sub-beamformer is obtained as the eigenvectors of the eigenvalue problem
\begin{equation}
[{\bf b}_1^{(g)} , {\bf b}_2^{(g)}, ...,{\bf b}_{D_g}^{(g)}] = \operatorname{eigs}({\bf R}_{sub,sum}^{(g)},{\bf R}_{sub,sum},D_g), \label{sub_geb}
\end{equation}
where ${\bf{R}}_{sub,sum}^{(g)} \triangleq  \frac{E_s}{G} \sum_{l=0}^{L_g-1}{\bf{R}}^{(g)}_{sub,l}$ and ${\bf{R}}_{sub,sum} \triangleq \frac{E_s}{G}\sum_{g=1}^{G} \sum_{l=0}^{L_g-1}{\bf{R}}^{(g)}_{sub,l} + N_o \bf I$, where ${\bf R}_{sub,l}^{(g)} \triangleq \sum_{u=1}^{U_g} {\bf R}_{sub,l}^{(g_u)}$, and $ {\bf{R}}^{(g_u)}_{sub,l} \in \mathbb{C}^{N_s \times N_s } $ is the sub-CCM of $l^{th}$ MPC of user $g_u$ in the reduced dimension, which is obtained by using an array of size $N_s$ placed at the origin. This yields a phase offset in the effective channel matrix; however, digital precoding makes necessary phase correction in digital domain. %Besides, this approach assumes that users in a group are distinguishable but close in angular domain so that they can be grouped together. %Therefore, individual CCM's can be obtained for each user in a group.

\section{Compensation Methods for PA Nonlinearities} \label{CompMeth}

Two methods will be presented in order to compensate the effects of PA nonlinearities. First method is the well known DPD, which aims to predict the distortion caused by the nonlinear PA and predistorts the signal such that overall transmitted signal has the desired form. The other compensation method is based on Bussgang decomposition for each subcarrier at the user terminal. This method corresponds to frequecy domain equalization for the linear channel formed by the memory of PA, which is not considered in design of the digital precoder.

\subsection{DPD for Hybrid Beamforming Architecture}

In this section, we present the DPD design framework for hybrid beamforming architecture, which employs anti-beamforming in order to reduce the dimension of the observation space to the number of RF chains. In literature, there are several methods that address anti-beamforming, which is implemented by reverse phase shifting; however, these works consider phase only subarray networks as analog beamformer. However, this approach is not valid for the proposed GEB since beamforming gains are not unity in general and required anti-beamforming is more complicated. Consider the transmitted signal, $\mathit{{\bf{y}}_n} = [y_{n}^{(1)} , y_{n}^{(2)}, ... , y^{(N_t)}_n]^T $, which can be expressed in terms of MP basis functions as ${\bf{y}}_n = \sum_{\varpi=-\Pi+1}^{\Pi-1}  \sum_{\upsilon =0}^{\Upsilon-1} \alpha_{\varpi,\upsilon} {\bf \bar{x}}_{n-\varpi}   \odot |{\bf \bar{x}}_{n-\varpi}|^{2\upsilon}$, where $ {\bf{\bar{x}}}_n $ is the analog beamformed signal. By expressing $ {\bf{\bar{x}}}_n $ in terms of digital predistorted signal, one can obtain
%\begin{equation}
%y_{n}^{(m)} = \Psi(\bar{x}^{(m)}_n). \label{tx_signal}
%\end{equation}

% In \eqref{tx_signal}, $ \Psi(\cdot)$ represents, PA nonlinearity defined in \eqref{pa_output}. Thus, transmitted signal from the antenna array can be expressed as,

%\begin{equation}
%{\bf{y}}_n = \sum_{\varpi=-\Pi+1}^{\Pi-1}  \sum_{\upsilon =0}^{\Upsilon-1} \alpha_{\varpi,\upsilon} {\bf \bar{x}}_{n-\varpi}   \odot |{\bf \bar{x}}_{n-\varpi}|^{2\upsilon},   \label{txarr}
%\end{equation}
%

\begin{equation}
{\bf{y}}_n = \sum_{\varpi=-\Pi+1}^{\Pi-1}  \sum_{\upsilon =0}^{\Upsilon-1} \alpha_{\varpi,\upsilon} [({\bf B\hat{x}}_{n-\varpi}) \odot |{\bf B \hat{x}}_{n-\varpi}|^{2\upsilon}], \label{pa_y_sig}
\end{equation}
where $ \odot $ denotes the Hadamard product and for simplicity all PA's are assumed to be identical. In the proposed architecture, $ N_t \times 1 $ observation vector, $ {\bf y}_n$, is used to design $ D $ DPD units. Hence, it is necessary to project the $ N_t \times 1 $ observations to the reduced dimension of $ D \times 1 $. For this purpose, consider the beamformed signal, $ \mathit{\bf{\bar{x}}}_u = {\bf{B}}\mathit{\bf{\hat{x}}}_n $, where it can be written in terms of sub-beamformers ${\bf{\bar{x}}}_u = {\bf{B}}^{(1)}{\bf{\hat{x}}}^{(1)}+ {\bf{B}}^{(2)}{\bf{\hat{x}}}^{(2)} + \ldots + {\bf{B}}^{(G)}{\bf{\hat{x}}}^{(G)} \label{bf_sign}$.

%\begin{equation}
%%{\bf{\bar{x}}}_u = [{\bf{B}}^{(1)}, {\bf{B}}^{(2)}, ..., {\bf{B}}^{(G)}]\begin{bmatrix}
%        {\bf{\hat{x}}}^{(1)} \\
%         \vdots\\
%         {\bf{\hat{x}}}^{(G)}
%        \end{bmatrix}, \label{abf_sig}
%\end{equation}

%\begin{equation}
%	{\bf{\bar{x}}}_u = {\bf{B}}^{(1)}{\bf{\hat{x}}}^{(1)}+ {\bf{B}}^{(2)}{\bf{\hat{x}}}^{(2)} + \ldots + {\bf{B}}^{(G)}{\bf{\hat{x}}}^{(G)}. \label{bf_sign}
%\end{equation}

In this study, anti-beamforming is performed by projecting the observation vector onto the subspaces, which are spanned by the sub-beamformers. DPD's for each group are designed based on the observation vector projected onto corresponding sub-beamformer space. Proposed approach employs pseudo-inverses of the sub-beamformers for the projection via anti-beamforming matrix, $ {\bf B}_{ab} =  [ ([{\bf{B}}^{(1)}]^\#)^T,([{\bf{B}}^{(2)}]^\#)^T , ..., ([{\bf{B}}^{(G)}]^\#)^T ]^T $, where $[{\bf{B}}^{(g)}]^\# = \left([{\bf{B}}^{(g)}]^H {\bf{B}}^{(g)}\right)^{-1}[{\bf{B}}^{(g)}]^H$ is the pseudo inverse matrix of $ {\bf{B}}^{(g)} $. Then the projected observation vector, ${\bf \tilde{x}}_n = {\bf B}_{ab} {\bf{y}}_n$, becomes

%\begin{equation}
%{\bf B}_{ab} =  [ ([{\bf{B}}^{(1)}]^\#)^T,([{\bf{B}}^{(2)}]^\#)^T , ..., ([{\bf{B}}^{(G)}]^\#)^T ]^T \label{a_bf}
%\end{equation}
%

%
%\begin{fleqn}
\begin{equation}
\begin{split}
{\bf \tilde{x}}_n = \alpha_{0,0}  {\bf B}_{ab} {\bf B\hat{x}}_n +
 \sum_{\substack{\upsilon =1}}^{\Upsilon-1} \alpha_{0,\upsilon} {\bf B}_{ab} [({\bf B\hat{x}}_n) \odot |{\bf B \hat{x}}_{n}|^{2\upsilon}] +
\\
 \sum_{\substack{{\varpi=-\Pi+1,} \\ { \varpi \neq 0}}}^{\Pi-1}  \sum_{\substack{\upsilon =0}}^{\Upsilon-1} \alpha_{\varpi,\upsilon} {\bf B}_{ab} [({\bf B\hat{x}}_{n-\varpi}) \odot |{\bf B \hat{x}}_{n-\varpi}|^{2\upsilon}],
\label{a_bf1}
\end{split}
\end{equation}
%\end{fleqn}
%
where $ {\bf B}_{ab} {\bf B} \approx {\bf I}_D$ since sub-beamformers are designed to be nearly orthogonal to each other. Second and third terms in \eqref{a_bf1} correspond to remaining interference, which depends on the distorted signals on all RF chains due to nonlinearity. Power of the interference term limits the performances of DPD's since each DPD is designed by using a single RF chain. After obtaining the projected values, one can form the error signal as ${\bf e}_n = {\bf \hat{x}}_{n} - {\bf \tilde{x}}_n$. Error signal, ${\bf e}_n$, is used to design the polynomial DPD by using indirect learning architecture (ILA) for $ d^{th}$ RF branch
\begin{equation}
{{\hat{x}}}_n^{(d)} = \sum_{\varpi=-\Pi'+1}^{\Pi'-1}  \sum_{\upsilon =0}^{\Upsilon'-1} w^d_{\varpi,\upsilon} {x}^{(d)}_{n-\varpi} |x^{(d)}_{n-\varpi}|^{2\upsilon},   \label{dpd_exp}
\end{equation}
where, $ w^d_{\varpi,\upsilon}$ are the DPD coefficients and they can be found by using least squares (LS) method proposed in \cite{GMP} as

\begin{equation}
{\bf w}_b^d ={\bf w}_{b-1}^d + \beta ({ \bf X}_d^H { \bf X}_d)^{-1}{ \bf X}_d^H {\bf e}^{(d)}, \label{dpd_coeff}
\end{equation}
where $ { \bf w}_b^d $ is the DPD coefficient vector calculated at $b^{th}$ block, ${ \bf X}_d $ is the observation matrix whose elements are $ \tilde{x}^{(d)}_{n-\varpi} |\tilde{x}^{(d)}_{n-\varpi}|^{2\upsilon}$, $ {\bf e}^{(d)} \triangleq \left[{\bf e}_n[d] \; {\bf e}_{n-1}[d] \;...\; {\bf e}_{n-P+1}[d]\right]^T $ is the error vector for $d^{th}$ RF branch, where ${\bf e}_n[d]$ is the $d^{th}$ element of ${\bf e}_n$, $P$ is the number of samples, and $\beta$ is the step-size for the adaptation. In this study, memory and nonlinearity order for the DPD is chosen as $\Pi' = 4$ and $\Upsilon' = 4$, respectively.  %It should be noted that DPD for each RF chain is designed individually due to multiplexing caused by analog beamformer even if all PA's are identical.

\subsection{Post Equalization at User Terminal}

A linear post equalizer is proposed in order to compensate the memory effects of PA's on the transmit chain as a simpler alternative to DPD. At BS, a pre-equalization is applied by means of digital precoding based on CSI acquired by uplink training. However, there is an undesired linear channel due to memory of PA's, which is ignored during the design of digital precoder. Therefore, an equalization procedure is necessary at the receiver side. In this work, Bussgang decomposition is applied for each sub-carrier at the receiver. Received signal at $g_u^{th}$ user after FFT operation for each subcarrier, $r^{(g_u)}_k $, can be expressed as \cite{SER1}
\begin{equation}
r^{(g_u)}_k = \alpha_k^{(g_u)} d^{(g_u)}_k + \eta^{(g_u)}_k,
\end{equation}
where $\alpha_k^{(g_u)}$ is the frequecy domain Bussgang coefficient and $\eta^{(g_u)}_k$ is the distortion term. Since Bussgang coefficient is defined for each subcarrier, it can be interpreted as the frequency domain channel for the corresponding subcarrier as
\begin{equation}
\alpha_k^{(g_u)}= \frac{\mathbb{E}\{ r_k^{(g_u)}[d_k^{(g_u)}]^{*}\}}{\mathbb{E}\{ |d_k^{(g_u)}|^2 \}}. \label{buss_coeff_post}
\end{equation}
By using scalar channel coefficient given in \eqref{buss_coeff_post}, symbol estimate is obtained as
\begin{equation}
\hat{d}_k^{(g_u)} = \frac{r_k^{(g_u)}}{\alpha_k^{(g_u)}}.
\end{equation}

\section{Performance Measures}\label{PerfMeas}
In this study, performances of the beamforming structures are evaluated via several performance criteria. Firstly, power spectral density (PSD) of the received signal at different locations  are evaluated. Beampatterns for the in-band and out-of-band radiation are presented in order to understand the effect of hybrid beamforming on the radiation pattern for different architectures. Then, by using GMI metric, AIR's for the considered architectures are obtained via the mismatched capacity framework. Lastly, a comprehensive analysis on BER performance is pursued by deriving signal to interference plus noise ratio (SINR) expressions for different array architectures.

\subsection{Radiation Patterns}

Radiation pattern is evaluated by investigating the PSD of the recieved signal at different angles. For this purpose, channel for any direction, $(\theta)$, is approximated as the steering vector, ${\bf a}^H(\theta)$, pointing that direction.
%\begin{equation}
%r^\theta_n = {\bf a}^H(\theta) {\bf y}_n. \label{rx_signal_theta}
%\end{equation}
Then PSD,  $S_r^\theta(f) = \sum_{t = - \infty}^{\infty} R^\theta_t e^{-j2 \pi f t}$, is obtained by taking the Fourier Transform of the autocorrelation function, $R^\theta_t = {\mathbb{E} } [r_n^\theta(r_{n-t}^\theta)^H]$, of the artificially generated received signal, $r^\theta_n = {\bf a}^H(\theta) {\bf y}_n$. Using $S_r^\theta(f) $, in-band and out-of-band radiation powers are calculated as

%\begin{equation}
%S_r^\theta(f) = \sum_{t = - \infty}^{\infty} R^\theta_t e^{-j2 \pi f t} \label{PSD}
%\end{equation}
%

\begin{equation}
P_{ib}(\theta) = \int_{-B/2}^{B/2} S_r^\theta(f) df, \quad P_{ob}(\theta) = max\left \{\int_{-3B/2}^{-B/2} S_r^\theta(f) df, \int_{B/2}^{3B/2} S_r^\theta(f) df \right\},\label{ib}
\end{equation}
where $ P_{ib}(\theta)$ and $ P_{ob}(\theta)$ are the in-band and OOB radiation patterns in $\theta $ direction respectively, and $B = \frac{2\pi}{\mu}$ is the normalized signal bandwidth.	

\subsection{Generalized Mutual Information (GMI)}

In this section, mismatched GMI concept, which is used to obtain a lower bound for the achievable rate for the channels with unknown  probability density function (PDF), is presented. In the considered problem, PA can be considered as a nonlinear channel, whose PDF is not known. In \cite{Capacity:2}, constrained capacity for the channels with known PDF is expressed by using GMI as

\begin{equation}
C = \log_2M - E_{d,r}\left[\log_2\left(\frac{\sum_{d'\in A_d} p(r|d')}{p(r|d)}\right)\right],
\end{equation}
where $A_d$ is QAM symbol alphabet, $M$ is the modulation order. $p(r|d)$ is the conditional PDF of the decoded signal $ r$ given that symbol $d$ is sent. In AWGN channel case, $p(r|d)$ has the Gaussian form; however, due to the nonlinearity caused by PA, Gaussian PDF is not valid. Thus, in this study, we employ an approximation of this PDF so that mismatched capacity introduced in\cite{GMI:1,GMI:2,Zglgn}, can be employed. A lower bound on the mismatched decoding capacity, where assumed PDF is utilized, is expressed as

\begin{equation}
C_M = \log_2M - E_{d,r}\left[\log_2\left(\frac{\sum_{d'\in A_d} \tilde {p}(r|d')}{\tilde {p}(r|d)}\right)\right], \label{capacity}
\end{equation}
where $ \tilde {p}(r|d') $ is the mismatched PDF.

%\subsubsection{Mismatch PDF at the receiver side}
In order to write the mismatched PDF, one should relate the decoded symbol to actual symbol. For this purpose, we use Bussgang theorem in order to decompose the signal into desired and distortion terms in frequency domain as, $r_k = \hat{\alpha}_k d_k + \eta_k$, where $\hat{\alpha}_k$ is the Bussgang coefficient per subcarrier and $ \eta_k$ is the distortion term with power $ \sigma^2_{\eta_k} = \mathbb{E}[\eta_k^H\eta_k]$ and it is orthogonal to linear signal term, $\mathbb{E}\left[d_k^H\eta_k\right]=0$. Bussgang coefficient, $\hat{\alpha}_k$ and the distortion variance, $\sigma^2_{\eta_k}$ for each subcarrier $k=0,1,\ldots,K-1$ can be found by Wiener filtering
%
%\begin{equation}
%r_k = \hat{\alpha}_k d_k + \eta_k, \label{buss}
%\end{equation}
%

\begin{equation}
\hat{\alpha}_k = \frac{\mathbb{E}[d_k^Hr_k]}{\mathbb{E}[d_k^Hd_k]},  \qquad
\sigma^2_{\eta_k} = \mathbb{E}[|r_k - \hat{\alpha}_k d_k|^2],
\label{buss_coef}
\end{equation}
and mismatched PDF is expressed in terms of Bussgang coefficient and the distortion power as

\begin{equation}
\tilde {p}(r_k|d_k)  = \frac{1}{\pi \sigma^2_{\eta_k}} \operatorname{exp}\left( - \frac{|r_k-{\hat{\alpha}}_k d_k|^2}{\sigma^2_{\eta_k}} \right), \label{PDFpost}
\end{equation}
which is valid for systems that utilizes equalization at UT's. Also, mismatch PDF can also be written for the systems without any receiver processing by replacing ${\hat{\alpha}}_k$ with $\hat \alpha$, which is a single phase and amplitude correction term defined as
\begin{equation}
\hat{\alpha} = \frac{\sum_{k=0}^{K-1}r_k d_k^*}{\sum_{k=0}^{K-1}|d_k|^2}. \label{buss_coef1}
\end{equation}
Bussgang coefficient defined in \eqref{buss_coef1} can also be used for the systems employing DPD at the transmitter since DPD also compensates the memory effects of the PA's. Hence, equalization at UT is not necessary. Finally, achievable capacity bound can be obtained by inserting mismatch PDF expression \eqref{PDFpost} into \eqref{capacity} after averaging over subcarriers. In \eqref{buss_coef} and \eqref{buss_coef1}, Bussgang coefficients are considered as independent of constellation point due to multiplexing of different symbols in precoders.

\subsection{Bit-Error-Rate Analysis}

A comprehensive analysis on BER performances of the different array structures is carried out to obtain an analytical approximation for probability of bit error. A general Bussgang decomposition is developed for multidimensional signal domain in which Bussgang coefficient has a matrix form rather than a single scalar due to correlation between signals on different antennas. Based on proposed Bussgang decomposition, SINR expression is obtained and BER is expressed as a function of SINR. For this purpose, nonlinear channel is modelled in frequency domain for different array architectures. For the rest of the analysis, digitally precoded, ${\bf x}_k^f$, and analog beamformed, ${\bf y}^f_k$, signals are defined in frequency domain as

\begin{equation}
{\bf x}_k^f = \frac{1}{\sqrt{\mu K}}\sum_{n=0}^{\mu K-1} {\bf x}_n e^{-j\frac{2\pi}{\mu K}nk }, \qquad
%\end{equation}
%\begin{equation}
{\bf y}_k^f = \frac{1}{\sqrt{\mu K}}\sum_{n=0}^{\mu K-1} {\bf y}_n e^{-j\frac{2\pi}{\mu K}nk }.
\end{equation}

\subsubsection{Fully Connected Array}

Conventional approaches consider Bussgang decomposition on PA basis; however, in this work, we propose a new decomposition scheme where analog beamformer is also taken into account together with PA's. Consequently, a spatio-frequency Bussgang decomposition is performed which is a mapping from reduced dimensional digital precoded signal vector, ${\bf x}_k^f$, to full dimensional transmitted signal, ${\bf y}_k^f$. Motivated by this approach, Bussgang decomposition is applied in matrix-vector form for each frequency bin, $k$, as
\begin{equation}
{\bf y}^f_k = {\bf A}_k{\bf x}_k^f + {\boldsymbol{\eta}}_k, \label{fully_buss}
\end{equation}
where ${\bf A}_k \in \mathbb{C}^{N\times D}$ can be considered as the linearized analog beamforming matrix after nonlinear amplification and $ {\boldsymbol{\eta}}_k \in \mathbb{C}^{N_t \times 1}$ is the nonlinear distortion vector for $k=0,1,...,\mu K-1$. In addition, ${\bf A}_k$ can be written in terms of sub-beamformers as ${\bf A}_k = [{\bf A}_k^{(1)},{\bf A}_k^{(2)},...,{\bf A}_k^{(G)}]$, which will be used later while defining different interference terms. ${\bf A}_k$ can be obtained as the generalization of the scalar Bussgang decomposition to matrix form

\begin{equation}
{\bf A}_k = \mathbb{E}\left[{\bf y}^f_k({\bf x}_k^f )^H\right]\left(\mathbb{E}\left[{\bf x}^f_k({\bf x}_k^f )^H\right]\right)^{-1}.
\end{equation}
One should note that if all the PA's were linear then Bussgang matrix would be analog beamformer matrix, ${\bf A}_k = \bf B$.
Due to correlation between different antennas, distortion terms are also correlated. Hence autocorrelation matrix for the distortion vector, ${\bf R}_{\boldsymbol{\eta}}^k$, is computed as
\begin{equation}
{\bf R}^k_{\boldsymbol{\eta}} \triangleq \mathbb{E} [{\boldsymbol{\eta}}_k {\boldsymbol{\eta}}_k^H]=  \mathbb{E}\left[  \left({\bf y}^f_k  -{\bf A}_k{\bf x}_k^f \right) \left({\bf y}^f_k  -{\bf A}_k{\bf x}_k^f \right)^H \right]. \label{acm}
\end{equation}

\subsubsection{Partially Connected Array}
In partially connected array structure, Bussgang coefficients are defined as vectors for each sub-array since each sub-array is driven by a single RF chain. Hence, transmitted signal for the sub-array of $g_d^{th}$ RF chain can be expressed as
\begin{equation}
{\bf y}^{(g_d),f}_k = {\bf a}^{(g_d)}_k{ x}_k^{(g_d),f} + {\boldsymbol{ \eta}}_k^{(g_d)}, \quad d = 1,\ldots, D_g\quad g = 1,\ldots,G \label{sub_sig}
\end{equation}
where $g_d$ is the $d^{th}$ RF chain for group $g$. Then linearized sub-beamformer can be expressed as
\begin{equation}
{\bf A}_k^{(g)} = {\bf e}_{G,g} \otimes \sum_{d=1}^{D_g} {\bf E}_{D_g,d} \otimes  {\bf a}^{(g_d)}_k, \label{subarraybuss}
\end{equation}
where $\otimes$ denotes Kronecker product, ${\bf e}_{G,g} $ is the $G$ dimensional elementary vector whose elements are $0$ except $g^{th}$ entry which is $1$, and ${\bf E}_{D_g,d} $ is the $D_g$ dimensional elementary matrix whose elements are $0$ except $d^{th}$ diagonal entry which is $1$. By using signal model given in \eqref{sub_sig}, one can define Bussgang vector for partially connected array as

\begin{equation}
 {\bf a}^{(g_d)}_k = \frac{ \mathbb{E}\left[{\bf y}^{(g_d),f}_k\left({x}_k^{(g_d),f} \right)^*\right]}{\mathbb{E}[|x_k^{(g_d),f}|^2]},
\end{equation}
and distortion covariance matrix, ${\bf R}^k_{\boldsymbol{\eta}} = \operatorname{bdiag}\left \{ \operatorname{bdiag} \left \{ {\bf R}^{(g_d),k}_{\boldsymbol{\eta}} \right \}_{d=1}^{D_g} \right\}_{g=1}^G$, has the block diagonal form since only the distortion terms in the same subarray are correlated, where ${\bf R}^{(g_d),k}_{\boldsymbol{\eta}} \triangleq \mathbb{E}\left[{{\boldsymbol{\eta}}}^{(g_d)}_k \left({{\boldsymbol{\eta}}}^{(g_d)}_k \right)^H\right] = \mathbb{E} \left[ ({\bf y}^{(g_d),f}_k - {\bf a}^{(g_d)}_k{ x}_k^{(g_d),f}) ({\bf y}^{(g_d),f}_k - {\bf a}^{(g_d)}_k{ x}_k^{(g_d),f})^H\right] $ is the autocorrelation matrix of the distortion terms of ${g_d}^{th}$ subarray.
%\begin{equation}
%{\bf R}^k_{\eta} = bdiag\{ {\bf R}^{(d),k}_{\eta} \}_{d=1}^D
%\end{equation}
%

\subsubsection{Fully Digital Beamforming}
For fully digital beamforming architectures, the standard Bussgang decomposition can be applied for each antenna in the system

\begin{equation}
{y}^{m,f}_k = {a}^{(m)}_k{ x}_k^{(m),f} + {\eta}^{(m)}_k, \quad k = 0,\ldots, \mu K-1, \label{dig_sig}
\end{equation}
where ${a}^{(m)}_k $ is the Bussgang coefficient for $m^{th}$ antenna defined as in \eqref{buss_coef}. %contrary to single Bussgang coefficient defined for each antenna for fully digital systems in literature \cite{abc,SER2,AIR1:1}. 

Having introduced the spatio-frequency Bussgang decomposition in multidimensional domain, equivalent received signal model at UT's can be obtained. Firstly, consider the input to nonlinear channel before the analog beamformer
\begin{equation}
{\bf x}_n = \frac{1}{\sqrt{K}} \sum_{k =0}^{\mu K-1}  \operatorname{bdiag}\{ \sqrt{c^{(g)}} {\bf W}_k^{(g)} \}_{g=1}^G {\bf d}_k e^{j \frac{2 \pi}{\mu K}kn},
\end{equation}
and also define the channel of the user $g_u$ in frequency domain as the $u^{th} $ column of the channel matrix in frequency domain, $ {\boldsymbol{ \omega}}_k^{(g_u)} \triangleq {\boldsymbol{ \Omega}}_k^{(g)}[:,u]$ and digital precoder for the same user is $ {\bf w}_k^{(g_u)} \triangleq{\bf W}_k^{(g)}[:,u]$ for $u = 1,...,U_g$. After taking DFT at UT for user $g_u$, one can express the received signal in frequency domain by using \eqref{fully_buss} as
\begin{equation} \label{rec_UT}
\begin{split}
r_k^{(g_u),f} &= [{\boldsymbol{ \omega}}_k^{(g_u)}]^H{\bf y}_k^f+\nu_k^{(g_u),f}, \qquad  k=0,1,...,K-1 \\
&= [{\boldsymbol{ \omega}}_k^{(g_u)}]^H({\bf A}_k{\bf x}_k^f + {\boldsymbol {\eta}_k})+\nu_k^{(g_u),f},\\
&= [{\boldsymbol{ \omega}}_k^{(g_u)}]^H\left[\sum_{g=1}^G \sqrt{\frac{c^{(g)}}{K}}{\bf A}_k^{(g)}{\bf W}_k^{(g)}{\bf d}_k^{(g)}+{\boldsymbol {\eta}}_k\right]+\nu_k^{(g_u),f}.
\end{split}
\end{equation}

Received signal in \eqref{rec_UT} is decomposed such that each interference term can be measured as
\begin{equation}
\begin{split}
r_k^{(g_u)} &= \underbrace{\sqrt{{\frac{c^{(g)}}{K}}}[{\boldsymbol{ \omega}}_k^{(g_u)}]^H{\bf A}_k^{(g)}{\bf w}_k^{(g_u)}d_k^{(g_u)}}_\text{desired signal term}+ \underbrace{\sqrt{{\frac{c^{(g)}}{K}}}\sum_{\substack{u'=1 \\ u' \neq u}}^{U_g}[{\boldsymbol{ \omega}}_k^{(g_u)}]^H{\bf A}_k^{(g)}{\bf w}_k^{(g_{u'})}d_k^{(g_{u'})}}_\text{intra-group interference}\\
&+\underbrace{\sum_{\substack{g'=1 \\ g'\neq g}}^{G}\sqrt{{\frac{c^{(g')}}{K}}}[{\boldsymbol{ \omega}}_k^{(g_u)}]^H{\bf A}_k^{(g')}{\bf W}_k^{(g')}{\bf d}_k^{(g')}}_\text{inter-group interference}+ \underbrace{[{\boldsymbol{ \omega}}_k^{(g_u)}]^H{\boldsymbol{ \eta}}_k}_\text{nonlinear distortion} + \underbrace{\nu_k^{(g_u),f}}_\text{AWGN}. \label{rec_dec}
\end{split}
\end{equation}

By using \eqref{rec_dec}, average received SINR at the UT for a particular channel is obtained as
\begin{equation}
SINR^{(g_u)}_k = \frac{\mathbb{E}_{\bf H}\left[{\frac{c^{(g)}}{K}} \left| [{\boldsymbol{ \omega}}_k^{(g_u)}]^H{\bf A}_k^{(g)}{\bf w}_k^{(g_u)}\right|^2 \; \middle| \; {\bf h}^{(g_u)}\right]}{\mathbb{E}_{\bf H}\left[\sigma^2_{I,N,k} \; \middle| \;{\bf h}^{(g_u)}\right]}\label{SINR}
\end{equation}
where $\sigma^2_{I,N,k}$ is the total interference-plus-noise power expressed as
\begin{equation}
\begin{split}
\sigma^2_{I,N,k}  =&{ {{\frac{c^{(g)}}{K}}}\sum_{\substack{u'=1 \\ u' \neq u}}^{U_g}|[{\boldsymbol{ \omega}}_k^{(g_u)}]^H{\bf A}_k^{(g)}{\bf w}_k^{(g_{u'})}|^2}  +  \sum_{\substack{g'=1 \\ g'\neq g}}^{G}{{\frac{c^{(g')}}{K}}}[{\boldsymbol{ \omega}}_k^{(g_u)}]^H{\bf A}_k^{(g')}{\bf W}_k^{(g')}[{\bf W}_k^{(g')}]^H  [{\bf A}_k^{(g')}]^H{\boldsymbol{ \omega}}_k^{(g_u)}\\
& + [{\boldsymbol{ \omega}}_k^{(g_u)}]^H {\bf R}_{{\boldsymbol{\eta}}}^k{\boldsymbol{ \omega}}_k^{(g_u)}  + N_o.
\end{split} 
\end{equation}
By using SINR expression given in \eqref{SINR}, the average BER can be obtained as
\begin{equation}
P_b \approx \frac{1}{UK} \sum_{g=1}^{G} \sum_{u=1}^{U_g} \sum_{k=0}^{K-1} \frac{4}{\log_2(M)}\left( 1- \frac{1}{\sqrt{M}} \right) \mathbb{E}_{{\bf h}^{(g_u)}} \left[ Q\left(\sqrt{\frac{3}{M-1}SINR^{(g_u)}_k} \right) \right],
\end{equation}
where $Q(x)=\frac{1}{\sqrt{2 \pi}} \int_x^{\infty} e^{(-t^2/2)}dt$.

\section{Numerical Results} \label{NumRes}

Numerical results are presented to compare different beamforming architectures under PA nonlinearities in terms of their radiation patterns, AIR and BER performances. Furthermore, analytical expression of BER is verified by the Monte Carlo trials. In this study, all PA's are assumed to be identical but their operating points depend on the gain of analog beamforming. In the simulations, PA model for $\sim 2$ GHz commercially available GaAs PA, which is presented in \cite{WEBSITE:1}, is used. $N_t = 96$ antennas having ULA geometry is placed at BS. There are $D = 6$ RF chains so each sub-array consists $N_s = 16$ antennas in partially connected array structure. In the considered scenario, there are $G=3$ groups, each group having $2$ users and also there is a victim user who is served by another BS, whose CCM is computed and involved in ${\bf R}_{sum}$ to place deep null in that direction. The user and victim distribution is summarized in Table  \ref{table:scenario}. It is assumed that a strong MPC always exists for each angular sector with Rician factor $\left|{{\kappa^{(g_u)}_l}}\right|^2/{{\gamma_l^{(g_u)}}}=10$. Also, the power of the first MPC is $3 \; dB$ larger than the power of second MPC and they are seperated by $2$ symbols duration. In generated OFDM signal, number of subcarriers and IFFT size are $K =550$  and $\mu K = 4096$, respectively, and cyclic prefix length is selected as $N_{cp}=20$.

\begin{figure}
\centering
    \includegraphics[scale=0.75]{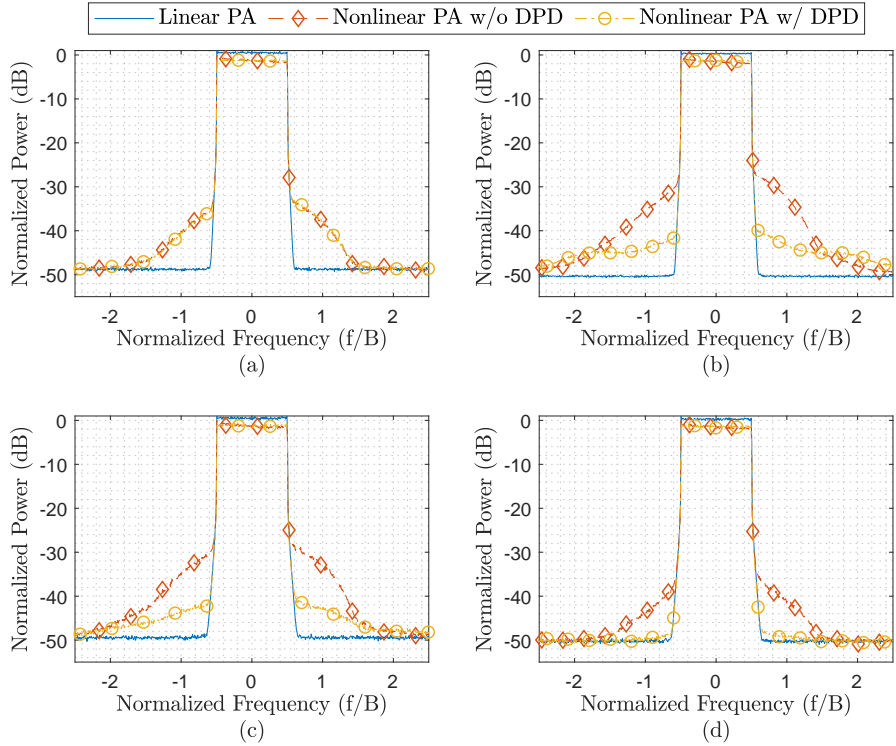}
    \caption{Normalized average power spectral density of intended users for different array architrectures (a) fully connected (b) partially connected with GEB (c) partially connected with phase only subarray (d) fully digital beamforming.}
    \label{AvgPSD}
\end{figure}

\begin{table}
\caption{Scenario}
\begin{center}
\begin{tabular}{|c|c|c|c|c|}

\hline
Group  &  \shortstack{ AS of $1^{st}$ MPC of \\ \vspace{0.5mm} \\ User $1$} & \shortstack{ AS of $1^{st}$ MPC of \\ \vspace{0.5mm} \\ User $2$} & \shortstack{ AS of $2^{nd}$ MPC of \\ \vspace{0.5mm} \\ User $1$} &\shortstack{ AS of $2^{nd}$ MPC of \\ \vspace{0.5mm} \\  User $2$} \\
\hline
1 &  ${[-28^\circ , -25^\circ]}$ & ${[-25^\circ , -22^\circ]}$  & ${[-17^\circ , -14^\circ]}$  & ${[-14^\circ , -11^\circ]}$ \\
\hline
2 &  ${[-4^\circ , -1^\circ]}$ & ${[-1^\circ , 2^\circ]}$ &  ${[8.5^\circ , 11.5^\circ]}$ &${[11.5 ^\circ, 14^\circ]}$ \\
\hline
3 &   ${[24^\circ , 27^\circ]}$& ${[21^\circ , 24^\circ]}$ & - & -\\
\hline
Victim &  ${[-39^\circ , -36^\circ]}$ & - & - & -\\
\hline
\end{tabular}
\label{table:scenario}
\end{center}
\end{table}
\emph{Power Spectral Density and Radiation Pattern Analysis:}
Power spectra and radiation patterns of the array structures are obtained in order to assess the severity of the unwanted power emissions in both frequency and spatial domains. In Fig. \ref{AvgPSD}, average PSD's of signals received by the users are shown. It is observed that OOB is a significant problem for all beamforming architectures. However, classical DPD is shown to be a feasible solution for investigated array structures, except fully connected array, since power leakage to adjacent channel is suppressed notably. In fully connected structure, however, classical DPD does not provide any improvement when OOB radiation is considered. It can be observed that DPD provides the best performance for fully digital system since there is a single DPD unit dedicated to each antenna element.

\begin{figure}
\centering
    \includegraphics[scale=0.75]{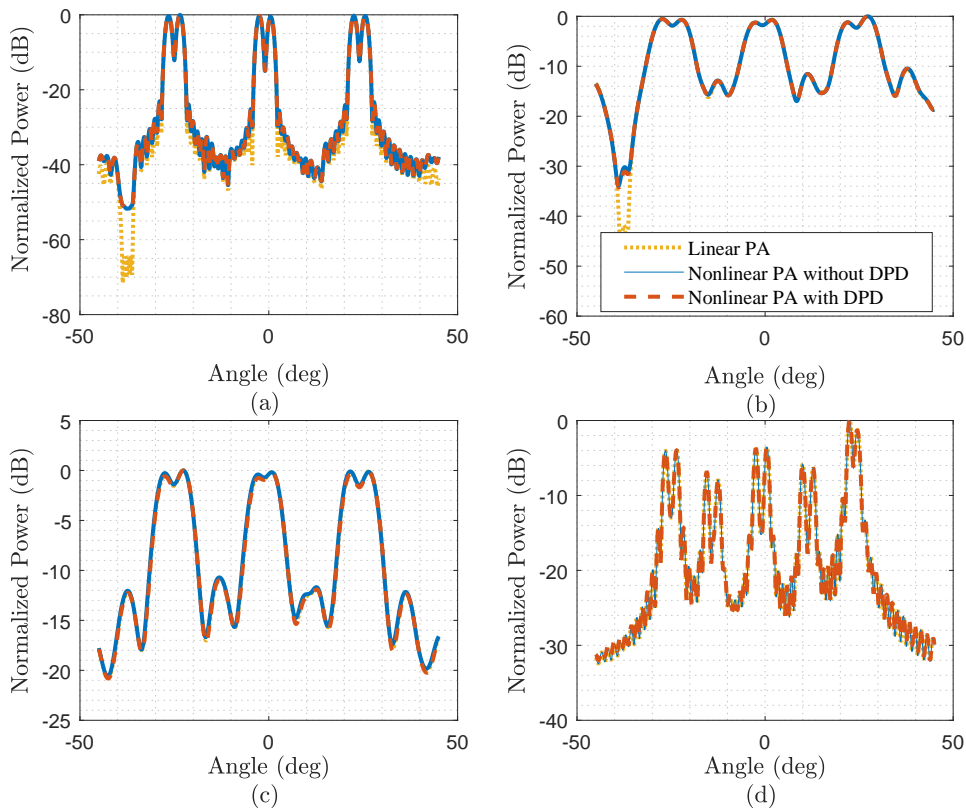}
    \caption{ In-band power radiation pattern from BS for  the interval of -50 and 50 degrees for different array architrectures (a) fully connected (b) partially connected with GEB (c) partially connected with phase only subarray (d) fully digital beamforming. Maximum in-band power is scaled to be 0 dB.}
    \label{ib_pattern}
\end{figure}

Furthermore, we investigate spatial distribution of in-band and out-of-band power emissions for different array architectures. Power angular spectrum for in-band radiation in Fig. \ref{ib_pattern}, shows that nonlinear effects do not distort the beampattern of the antenna array. In addition it can be observed that due to nonlinear distortion, null level of the array gets higher, which degrades the inter-group interference suppression capability. However, despite the higher null level, a significant amount of suppression is still achieved compared to phase only subarray.

\begin{figure}
\centering
    \includegraphics[scale=0.75]{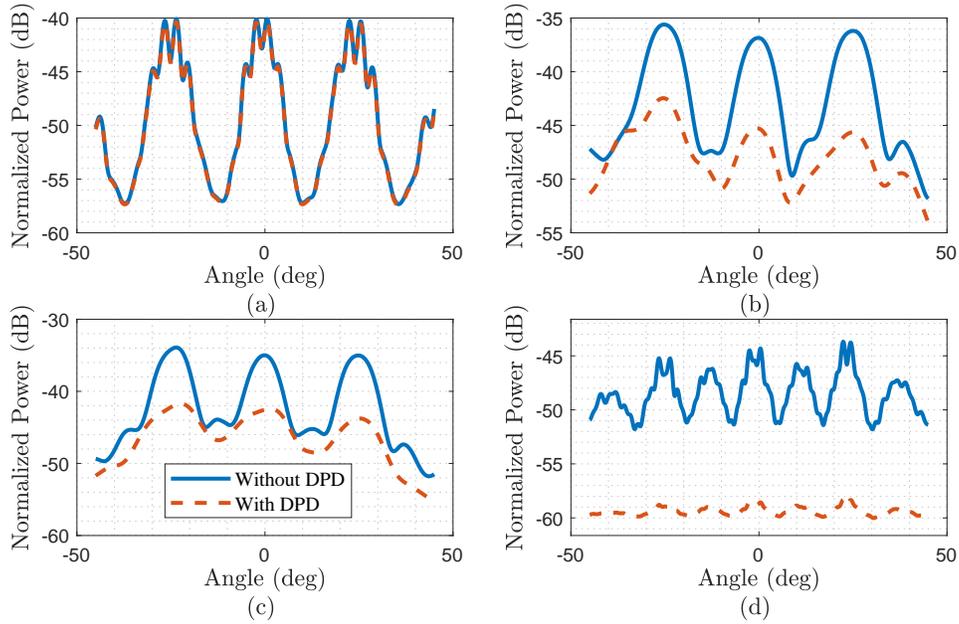}
    \caption{ Out-of-band power radiation pattern from BS for the interval of -50 and 50 degrees for different array architrectures (a) fully connected (b) partially connected with GEB (c) partially connected with phase only subarray (d) fully digital beamforming. Maximum in-band power is scaled to be 0 dB.}
    \label{oob_pattern}
\end{figure}

In Fig. \ref{oob_pattern}, power angular spectrum for out-of-band radiation is shown. Out-of-band band power is also concentrated in the served user directions. Similar to PSD analysis, DPD does not provide improvement in the OOB radiation in any direction as can be seen in Fig. \ref{oob_pattern}(a), for fully connected architecture. On the other hand, employing DPD at the transmitter significantly decreases OOB emission for partially connected array and fully digital precoding architectures. However, it is also observed that radiated power in fully connected architecture is not much higher than that of partially connected arrays with DPD since its multiplexing gain is larger than the multiplexing gain of partially connected array. The reason is that in fully connected structure all, $ N_t= 96$, are used to transmit the signal to a single user. On the other hand, only a sub-array of size, $N_s = 16$, is dedicated for a single user in partially connected systems. Also, at the UT, desired signal term is combined coherently, while distortion term is combined non-coherently, consequently, effects of the nonlinear distortion is averaged out in fully connected and fully digital architectures compared to partially connected architecture.

\begin{figure}
\centering
    \includegraphics[scale=0.6]{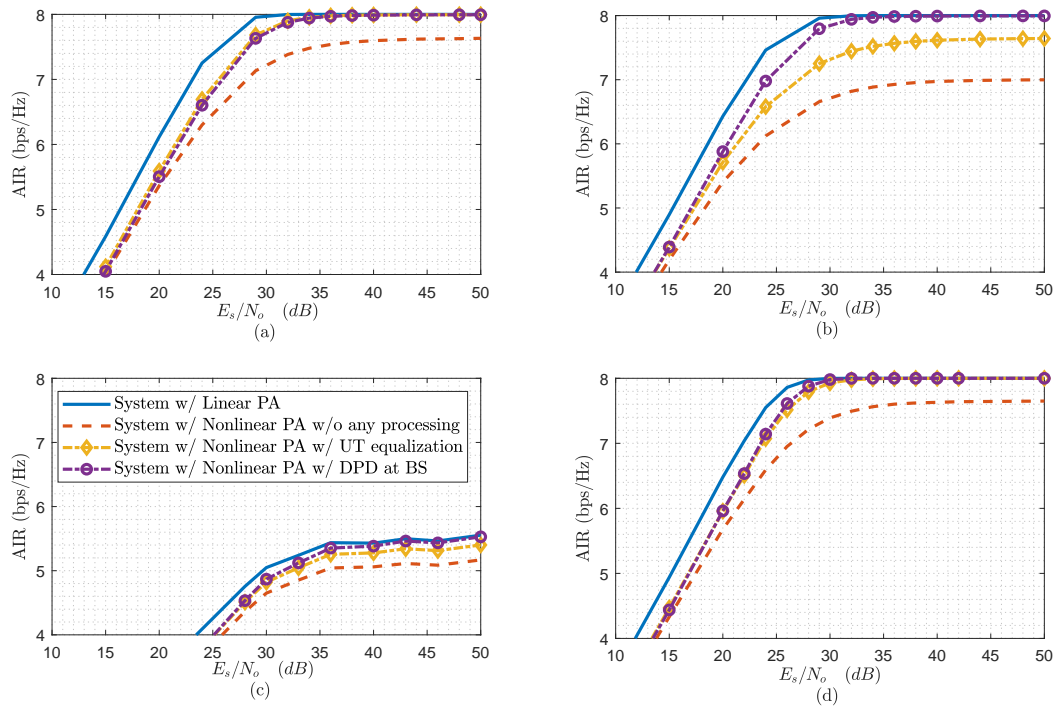}
    \caption{Average mismatched capacity of users for $256$ QAM for different array architectures (a) fully connected (b) partially connected with GEB (c) partially connected with phase only subarray (d) fully digital beamforming.}
    \label{GMI}
\end{figure}

\emph{AIR Analysis:}
Having investigated the spatial radiation of the transmitted power, we can now compare the AIR performances of the different beamforming architectures with different processing techniques. For this purpose, GMI analysis is carried out for these structures. Capacity curves are obtained for a constant transmitted power, $E_s$ and $E_s/N_o$, which is the received signal to noise ratio (SNR) per-subcarrier, is adjusted by modifying the receiver noise power. Firstly, from Fig. \ref{GMI}, it can be inferred that  array structures has vital impact on the capacity of the system. Phase only subarray cannot achieve the maximum AIR regardless of the noise level and nonlinear distortion. Even for linear PA, capacity of this architecture is limited due to inter-group interference. However, the other structures can successfully suppress both inter-group and intra-group interferences, which make them superior compared to phase only beamforming. Apart from array structure, nonlinear distortion introduced by PA has notable effect on the performance since systems with linear PA's can reach higher spectral efficiency for the same noise level compared to systems with nonlinaer PA's, until it is limited by the constellation order.

Both systems with fully connected and fully digital array architectures can achieve maximum capacity by employing either equalization at UT or DPD, and these methods exhibit close performances for both architectures. The reason is that nonlinear distortion is already suppressed by the diversity; therefore, linear processing is sufficient. On the other hand, multiplexing gain cannot be fully exploited by partially connected systems due to decrease in the effective antenna per user, which decreases the multiplexing gain. Therefore, maximum capacity cannot be reached by employing linear receiver processing in GEB subarray whereas it is achieved by conventional DPD since it suppresses nonlinear distortion as well. Consequently, classical DPD technique is proven to be a feasible solution. Lastly, one can see that the systems, which do not employ any processing, suffer from significant performance degradation for all array architectures.

\begin{figure}
\centering
    \includegraphics[scale=0.35]{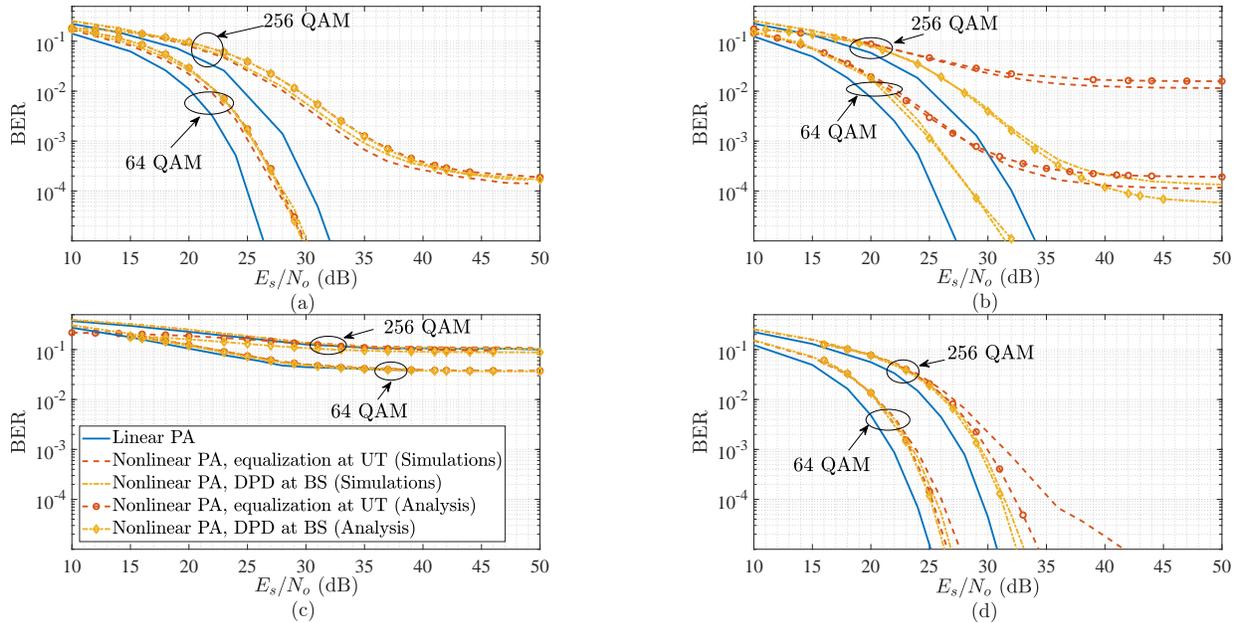}
    \caption{Analytical and simulation results for BER obtained for different array architectures (a) fully connected (b) partially connected with GEB (c) partially connected with phase only subarray (d) fully digital beamforming for $64$ and $256$ QAM}
    \label{BER}
\end{figure}

\emph{BER Analysis:}
In this section, analysis on uncoded BER performances of the array structures are evaluated. Derived analytical BER expression is verifed with  numerical Monte Carlo simulations for $64$ and $256$ QAM constellations.  Fig. \ref{BER} presents theoretical and  numerical BER curves. It can be seen that analytical BER curves are in compliance with the numerical results. However, there is a divergence in case of fully digital beamforming with $256$ QAM, which is acceptable since BER approximation for a higher order modulation is employed and which may not be exact but it sets a bound for BER. Also note that BER results for the system, which does not employ any processing, is not considered in BER analysis due to its poor AIR performance.

Firstly, it can be observed that phase only subarray exhibits significantly higher error floor compared to other architectures. For fully digital beamforming; on the other hand, both DPD and equalization methods perform close to linear PA in case of $64$ QAM; however, post-equalization method suffers from a performance degradation for $256$ QAM. 

Similar to fully digital structure, both DPD and equalization methods also perform close to linear PA for $64$ QAM for fully connected structure. However, there exists an error floor for the systems with nonlinear PA for $256$ QAM. From these results, one can also conclude that classical DPD does not provide any improvement for fully connected structure since in this structure, each PA is connected to all RF chains; however, each DPD is designed based on a single RF chain. Therefore, DPD's cannot compensate the nonlinear distortion, they only compensate the memory effects of PA's. This can be also verified by the simulations since equalization at UT and DPD at BS performs very closely.

In subarray with GEB, improvement in the performance becomes more apparent. For both $256$ and $64$ QAM, equalization at UT cannot prevent error floor. However, by employing DPD, error floors for $64$ and $256$ QAM are significantly reduced. Furthermore, error floor of subarray with GEB for $256$ QAM is slightly lower than that of fully connected architecture.

\section{Conclusions and Future Works} \label{conclds}

In this work, we presented a comprehensive performance analysis framework for evaluation of massive MIMO systems, which are subject to PA nonlinearities for different beamforming architectures.  In addition, anti-beamforming based indirect learning approach is adopted in design of DPD and a post-equalization method is proposed.  GEB based sub-array design is proposed in order to provide sufficient inter-group interference in analog domain, which cannot be achieved by conventional DFT beamformer. Both in-band and out-of-band radiation patterns are considered for different array architectures. Significant amount of OOB radiation results due to nonlinear distortion caused by PA's; however,  in band pattern is not significantly affected. Conventional DPD can achieve OOB reduction for all array architectures except fully connected architecture. DPD design for fully connected architecture that considers the connectivity of this architecture remains as a future work. GMI metric is employed in order to compare array architectures in terms of AIR. Observations on AIR performances are in compliance with that of OOB radiation. It can be observed that fully digital beamforming arcihtecture is robust agianst PA nonlinearities since maximum capacity can be achieved by using only linear equalization at the receiver. In addition, fully connected architecture provides acceptable performance under PA nonlinearities even if its OOB radiation cannot be recduced. On the other hand, DFT beamformer suffers from significant performance degradation even with linear PA due to poor performance of analog beamformer design. Lastly, analytical BER expressions are derived for different array architecutres based on spatio-frequency multidimensional Bussgang decomposition and BER performances of array architectures are compared. Sub-array with DFT beamformer exhibits poor performance while fully digital beamformer provides the best possible performance. Proposed sub-array with GEB outperforms fully connected architecture for higher order modulations; however, for lower order constellations fully connected architecture is more beneficial. As a result, sub-array with GEB provides tradeoff between OOB supression and BER performance. However, fully connected architecture would have promising performance with proper DPD design.

% or


%\appendices

%% you can choose not to have a title for an appendix
%% if you want by leaving the argument blank
%\section{}

%% use section* for acknowledgement
%\section*{Acknowledgment}


\ifCLASSOPTIONcaptionsoff
  \newpage
\fi



%\bibliographystyle{IEEEtranTCOM}
%\begin{thebibliography}{1}


\bibliography{references_bib} 
\bibliographystyle{IEEEtran}

%\end{thebibliography}

%\begin{biography}[{\includegraphics[width=1in,height=1.25in,clip,keepaspectratio]{mshell}}]{Michael Shell}


%\vfill


\end{document}